\documentclass[fleqn,12pt,twoside]{article}
\usepackage{espcrc1}

\usepackage{amsmath}
\usepackage{amssymb}

\usepackage{graphicx}

\newcommand{\be}{\begin{equation}}
\newcommand{\ee}{\end{equation}}

\newcommand{\chiPT}{$\chi$PT}

\newcommand{\HBchiPT}{HB$\chi$PT}

\newcommand{\mpi}{m_\pi}
\newcommand{\fpi}{f_\pi}
\newcommand{\fm}{\,{\rm fm}}
\newcommand{\mev}{\,{\rm MeV}}
\newcommand{\gev}{\,{\rm GeV}}

\title{
\vspace{-2.6cm}
\hfill \rm \null \hfill
\hbox{\normalsize ADP-05-02/T612} \\
\vspace{-2mm}
\hfill \hbox{\normalsize JLAB-THY-05-294} \\
\vspace{1.65cm}
Extrapolation of lattice QCD results beyond the power-counting regime}

\author{D.~B.~Leinweber\address[CSSM]{Special Research Center for the
                       Subatomic Structure of Matter, and		\\
                       Department of Physics, University of Adelaide
                       Adelaide SA 5005  Australia}%
                       \address[JLab]{Jefferson Lab, 12000
                       Jefferson Ave., Newport News, VA 23606, USA.},
        A.~W.~Thomas\addressmark[JLab],
        and 
        R.~D.~Young\addressmark[JLab]
}
       
\begin{document}

\maketitle

\begin{abstract}
Resummation of the chiral expansion is necessary to make accurate
contact with current lattice simulation results of full QCD.
Resummation techniques including relativistic formulations of chiral
effective field theory and finite-range regularization (FRR)
techniques are reviewed, with an emphasis on using lattice simulation
results to constrain the parameters of the chiral expansion.  We
illustrate how the chiral extrapolation problem has been solved and
use FRR techniques to identify the power-counting regime (PCR) of
chiral perturbation theory.  To fourth-order in the expansion at the
1\% tolerance level, we find $0 \le m_\pi \le 0.18$ GeV for the PCR,
extending only a small distance beyond the physical pion mass.
\end{abstract}

\section{INTRODUCTION}

Dynamical chiral symmetry breaking in QCD gives rise to an octet of
light mesons recognized as the (pseudo) Goldstone bosons of the
symmetry.  These light mesons couple strongly and give rise to a
quark-mass dependence of hadron observables which is nonanalytic in
the quark mass.  The Adelaide Group has played a leading role in
emphasizing the role of this physics in the chiral extrapolation of
lattice simulation results
\cite{Early,Moments,Young:2002cj,Young:2002ib,%
Leinweber:2003dg,Young:2004tb,Leinweber:2004tc}.
Most work in the baryon sector has focused on the nucleon and Delta
masses and the magnetic moments of octet baryons.

The established, model-independent approach to chiral effective field
theory is that of power counting, the foundation of chiral
perturbation theory (\chiPT).  However, this requires one to work in a
regime of pion mass where the next term in the truncated series
expansion makes a contribution that is negligible.  Given the emphasis
on determining the nucleon mass to 1\%, such neglected contributions
must be constrained to the fraction of a percent level.  As there is
no attempt to model the higher-order terms of the chiral expansion,
one simply obtains the wrong answer if one works outside this region.

There is now growing recognition that some form of resummation of the
chiral expansion is necessary in order to make contact with lattice
simulation results of full QCD, where the effects of dynamical
fermions are incorporated.  As we will demonstrate in the following,
the quark masses accessible with today's algorithms and supercomputers
lie well outside the regime of \chiPT\ in its standard form.  This
situation is unlikely to change significantly until it becomes
possible to directly simulate QCD on the lattice within twice the
squared physical mass of the pion and with suitably large lattice
volumes.  Within this range, knowledge of the first few terms of the
chiral expansion would be sufficient as higher-order terms of the
expansion are small simply because $m_\pi$ is small.  Straightforward
application of the truncated expansion of \chiPT\ would then be
possible.  However, one might wonder if the lattice techniques that
would allow simulations at light masses within $2 m_\pi^2$, might also
allow a calculation directly at the physical pion mass, obliterating
the chiral extrapolation problem altogether.  Unfortunately such
simulations are not possible in the foreseeable future, and one must
resort to more advanced techniques.

The resummation of the chiral expansion induced through the
introduction of a finite-range cutoff in the momentum-integrals of
meson-loop diagrams is perhaps the best known resummation method
\cite{Young:2002cj,Young:2002ib,Leinweber:2003dg,%
Young:2004tb,Leinweber:2004tc,Donoghue:1998bs,Borasoy:2002jv,Bernard:2003rp}
--- for alternative proposals, see also
Refs.~\cite{Djukanovic:2004px,Pascalutsa:2004je}.  However, there are
many ways of regularizing the loop integrals which lead to a
resummation of the expansion.  For example, relativistic formulations
lead to expressions for the nucleon self-energy which include
higher-order nonanalytic terms beyond the order one is calculating
\cite{Procura:2003ig}.  In some cases, the expressions have the
desirable property of becoming smooth and slowly varying as the pion
mass becomes moderately large \cite{Pascalutsa:2004ga}.  Indeed all
observables calculated on the lattice display a smooth slowly-varying
dependence on the quark mass, presenting a clear signal that higher
order terms of the chiral expansion sum to zero as the pion mass
becomes moderately large.

The ability to draw a curve through the lightest nucleon mass points
has led some to argue that it is essential to adopt a relativistic
approach in calculating the chiral expansion
\cite{Procura:2003ig,AliKhan:2003cu}.  However, one might have some
concern that a relativistic expression is so important in what is
supposed to be a {\em low-energy} effective field theory.  Indeed, a
Taylor expansion of the relativistic expression produces exactly the
same leading-nonanalytic (LNA) term as obtained in the heavy-baryon
formulation of chiral perturbation theory (\HBchiPT).  Upon including
the Delta in \HBchiPT\ the next-to-leading nonanalytic (NLNA)
structure, $\mpi^4\log\mpi$ is also obtained.

In the relativistic formulation, an $m_\pi^5$ correction is also
generated at one-loop order.  However, the rapid convergence of the
Taylor series generated by the full relativistic loop calculation
\cite{Procura:2003ig} emphasizes that the relativistic corrections are
small.  One should not assume that the higher-order terms of the
heavy-baryon expansion are included completely.  Rather a particular
resummation of the chiral expansion has been obtained which allows the
self-energy of the nucleon to evolve slowly as the pion mass grows,
thus mimicking the lattice simulation results.

\begin{figure}[t]
\begin{center}
\includegraphics[width=0.9\hsize]{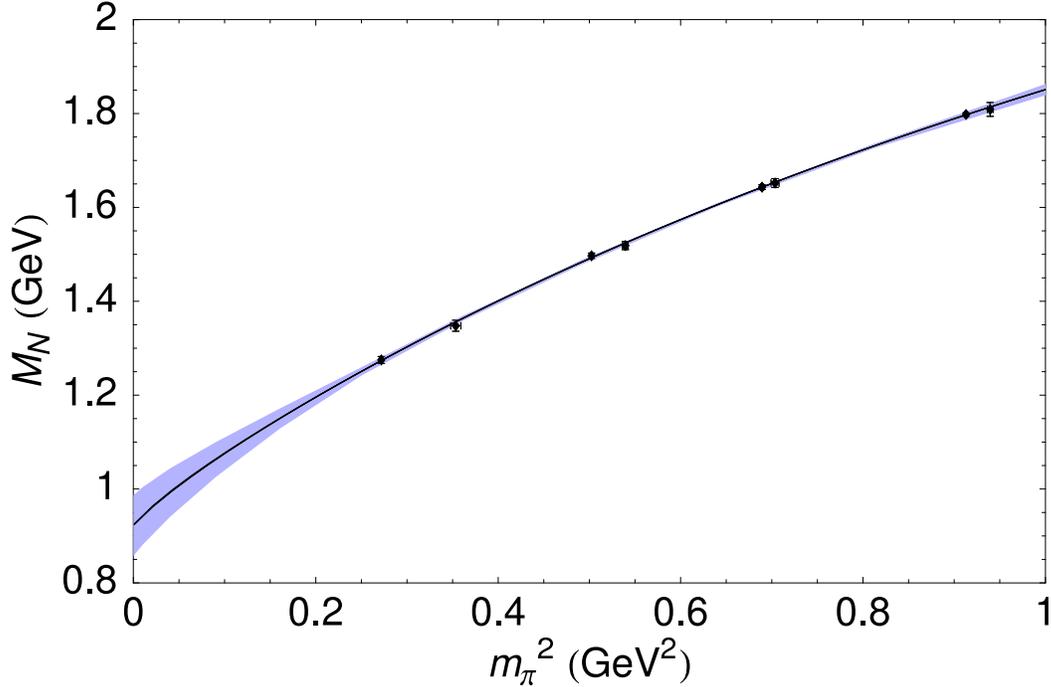}
\end{center}
\vspace{-1.3cm}
\caption{Extrapolation of CP-PACS collaboration simulation results
  \protect\cite{AliKhan:2001tx} to the chiral limit
  using finite-range regularization \cite{Leinweber:2003dg}.  Differences between the
  illustrated dipole, monopole, Gaussian and theta-function regulators
  cannot be resolved on this scale.  The one-standard deviation
  error bound for the dipole extrapolation is also illustrated.
\vspace{-0.3cm}
}
\label{FRR}
\end{figure}

Perhaps the most astonishing discovery in FRR chiral effective field
theory, is that the term-by-term details of the higher-order chiral
expansion are largely irrelevant in describing the chiral
extrapolation of simulation results.  Figure~\ref{FRR} displays the
extrapolation \cite{Leinweber:2003dg} of lattice simulation results of
full QCD from the CP-PACS collaboration \cite{AliKhan:2001tx}.  A
variety of finite-range regularizations are illustrated, including
dipole, monopole, Gaussian and theta-function regulators, as described
in greater detail below.  FRR chiral effective field theory is
mathematically equivalent to \chiPT\ to any finite order and in this
case we work to order $m_\pi^4 \log m_\pi$.

The curves are indistinguishable and produce physical nucleon masses
which differ by less than 0.1\%.  This is despite the fact that the
coefficients of the higher-order terms ($m_\pi^5$ and beyond)
appearing in the FRR expressions differ significantly.  For example,
the coefficient of the $m_\pi^5$ term vanishes for the theta-function
regulator, whereas the monopole, dipole and Gaussian regulators
provide $-44$, $-34$ and $-30\gev^{-4}$ respectively.  What these
expansions do have in common is that the loop-integrals vanish as the
quark masses grow large.  The contribution of any individual
higher-order term is largely irrelevant.  The only thing that really
counts is that there are other terms that enter to ensure the sum of
all terms of the loop integral approaches zero, in accord with what is
observed in lattice QCD calculations.  Of course, this beautiful
feature of FRR expansions would be lost if one were to truncate the
expansion at any finite order.  Resummation of chiral effective field
theory is essential to solving the chiral extrapolation problem.

In each of the aforementioned regulators, there is a single parameter
governing the range of the loop-momentum cutoff.  It is a parameter
which does not appear in standard chiral perturbation theory and as
outlined in detail below it plays no role in the terms of the FRR
expansion to the order one is working.  However, it does play a role
in determining the coefficients of the higher-order terms, beyond that
order.  As discussed in detail in
Ref.~\cite{Leinweber:2003dg}, it governs the convergence properties of
the chiral expansion.  Given the level of agreement between the curves
associated with different regulators, it is sufficient to parameterize
the remainder of the chiral expansion in terms of this single
parameter, $\Lambda$, and still maintain fraction of a percent
accuracy.

It is essential that $\Lambda$ remain finite.  In the limit $\Lambda
\to \infty$ one recovers the standard chiral expansion of \chiPT\
truncated at the next nonanalytic term beyond the order one is
working, spoiling the resummation of the expansion.  $\Lambda$ must be
constrained to values that ensure that the loop-momenta contributing
to loop integrals are relevant to the low-energy effective field
theory.  Indeed it is the erroneous contributions from large
loop-momenta beyond the realm of the low-energy effective field theory
that spoils the convergence properties of traditional \chiPT\
\cite{Donoghue:1998bs}.  The difficulty with this approach is that the
exact value of $\Lambda$ is not specified.  Because one must work to
finite order in the expansion, there is a residual dependence on
$\Lambda$.  The common approach is to vary $\Lambda$ over reasonable
values and include the observed variation in the final uncertainty
estimate.

\section{FORMAL BACKGROUND}
\label{renormalization}

In the usual formulation of effective field theory, the nucleon mass
as a function of the pion mass (given that $m_\pi^2 \propto m_q$)
has the formal expansion:
\begin{equation}
M_N = a_0 + a_2 m_\pi^2 + a_4 m_\pi^4 + a_6 m_\pi^6
      + \chi_\pi I_\pi + \chi_{\pi\Delta} I_{\pi\Delta} 
      + \chi_{\rm tad} I_{\rm tad}  \, .
\label{eq:formal}
\end{equation}
In practice, for current applications to lattice QCD, the parameters,
$a_i$, should be determined by fitting to the lattice results
themselves. The additional terms, $\chi_\pi I_\pi$, $\chi_{\pi\Delta}
I_{\pi\Delta}$ and $\chi_{\rm tad} I_{\rm tad}$, are loop corrections
involving the (Goldstone) pion.  As these terms involve the coupling
constants in the chiral limit, which are essentially model-independent
\cite{Li:1971vr}, the only additional complication they add is that
the ultra-violet behavior of the loop integrals must be regulated in
some way.

Traditionally, one uses dimensional regularization which (after
infinite renormalization of $a_0$ and $a_2$) leaves only the
non-analytic terms --- $c_{\rm LNA} m_\pi^3$ and $c_{\rm NLNA} m_\pi^4
\ln (m_\pi/\mu)$, respectively.  Within dimensional regularization one
then arrives at a truncated power series for the chiral expansion.  To
fourth order
\begin{equation}
M_N = c_0 + c_2 m_\pi^2 + c_{\rm LNA} m_\pi^3 + c_4 m_\pi^4 
      + c_{\rm NLNA} m_\pi^4 \log \frac{m_\pi}{\mu} + c_6' m_\pi^6 \, ,
\label{eq:drexp}
\end{equation}
where the bare parameters, $a_i$, have been replaced by the finite,
renormalized coefficients, $c_i$. Through the chiral logarithm one has
an additional mass scale, $\mu$, but the dependence on this is
eliminated by matching $c_4$ to ``data'' (in this case lattice
QCD). The prime on $c_6'$ denotes that this term is beyond the order
to which we are working, and therefore should not be expected to be
independent of regularization.

In line with the implicit $\mu$-dependence of the coefficients in the
familiar dimensionally regulated \chiPT , the systematic FRR expansion
of the nucleon mass is:
\begin{eqnarray}
M_N = a_0^\Lambda + a_2^\Lambda m_\pi^2 + a_4^\Lambda m_\pi^4 + a_6^\Lambda m_\pi^6
      + \chi_\pi I_\pi(m_\pi, \Lambda) + \chi_{\pi\Delta} I_{\pi\Delta}(m_\pi, \Lambda) 
      + \chi_{\rm tad} I_{\rm tad}(m_\pi, \Lambda)
\label{eq:finite}
\end{eqnarray}
where the dependence on the {\it shape} of the regulator is implicit.
The dependence on the value of $\Lambda$ and the choice of regulator
is eliminated, to the order of the series expansion, by fitting the
coefficients, $a_n^\Lambda$, to lattice QCD data. 

To illustrate how the $\Lambda$-dependence of the chiral expansion is
removed to the order one is working, we review the renormalization
procedure.  To leading one-loop order 
\begin{equation}
M_N = a_0 + a_2 \mpi^2 + \chi_\pi I_\pi\, ,
\label{eq:MNlead}
\end{equation}
where $\chi_\pi$ is the LNA coefficient of the nucleon mass expansion,
\begin{equation}
\chi_\pi = -\frac{3}{32\pi\fpi^2} g_A^2 \, ,
\label{eq:chiNpi}
\end{equation}
and $I_\pi$ denotes the relevant loop integral. In the heavy baryon
limit, this integral over pion momentum is given by
\begin{equation}
I_\pi = \frac{2}{\pi} \int_0^\infty dk\, \frac{k^4}{k^2+\mpi^2} \, .
\label{eq:INpi}
\end{equation}
This integral suffers from a cubic divergence for large momentum. The
infrared behavior of this integral gives the leading nonanalytic
correction to the nucleon mass. This arises from the pole in the pion
propagator at complex momentum $k=i\mpi$ and will be determined
independent of how the ultraviolet behavior of the integral is
treated. Rearranging Eq.~(\ref{eq:INpi}) we see that the pole contribution can
be isolated from the divergent part
\begin{equation}
I_\pi = \frac{2}{\pi} \int_0^\infty dk\, \left(k^2-\mpi^2\right) 
    + \frac{2}{\pi} \int_0^\infty dk\, \frac{\mpi^4}{k^2+\mpi^2} \, .
\label{eq:IpiPole}
\end{equation}
The final term converges and is given simply by
\begin{equation}
\frac{2}{\pi} \int_0^\infty dk\, \frac{\mpi^4}{k^2+\mpi^2} = \mpi^3 \, ,
\end{equation}
where we now recognize the choice of normalization of the loop
integral, defined such that the coefficient of the LNA term is set to
unity. This choice is purely convention and allows for a much
more transparent presentation of the differences in the chiral
expansion with various regularization schemes.

In the most basic form of renormalization we could simply imagine
absorbing the infinite contributions arising from the first term in
Eq.~(\ref{eq:IpiPole}) into a redefinition of the coefficients $a_0$ and
$a_2$ in Eq.~(\ref{eq:MNlead}). This solution is simply a minimal subtraction
scheme and the renormalized expansion can be given without making
reference to an explicit scale,
\begin{equation}
M_N = c_0 + c_2 \mpi^2 + \chi_\pi \mpi^3 \, ,
\label{eq:MSrenLNA}
\end{equation}
with the renormalized coefficients defined by
\begin{equation}
c_0 = a_0 + \chi_\pi \frac{2}{\pi} \int_0^\infty dk\, k^2 \, ,\quad
c_2 = a_2 - \chi_\pi \frac{2}{\pi} \int_0^\infty dk\, .
\end{equation}
Equation~(\ref{eq:MSrenLNA}) therefore encodes the complete quark mass
expansion of the nucleon mass to ${\cal O}(\mpi^3)$. This result will
be precisely equivalent to any form of minimal subtraction scheme,
where all the ultraviolet behavior is absorbed into the two
leading coefficients of the expansion. Such a minimal subtraction
scheme is characteristic of the commonly implemented dimensional
regularization.

This formulation has implied that the description of the pion-cloud
effects are equally well described at all momentum scales. As we know
that chiral effective field theory is only valid in the low momentum
regime, a minimal subtraction scheme will describe high momentum pion
modes which have no connection with physical reality.  This incorrect
short-distance physics must be removed through the analytic terms of
the expansion, spoiling the convergence of the expansion
\cite{Donoghue:1998bs}.

The pion-nucleon system in the real world is characterized by a
momentum scale associated with the size of the pion-cloud source.  The
axial form factor of the nucleon provides a radius of $\langle r^2
\rangle_{\rm axial}^{1/2}\simeq 0.66\fm$ \cite{Thomas:2001kw},
suggesting that the characteristic scale is of order $1/\langle r^2
\rangle_{\rm axial}^{1/2}\sim 0.3$--$0.4\gev$.  Hence one should
expect an improvement in the convergence properties of the chiral
expansion upon implementing a regularization which suppresses the
ultraviolet behavior of the loop integral.  We refer to any such
scheme as finite-range regularization (FRR).

It should be noted that a minimal subtraction scheme does offer the
advantage that all finite energy scales beyond the pseudo-Goldstone
boson mass are integrated out.  The finite size of the nucleon will
therefore emerge as the nucleon is dressed by higher order pion
cloud dressing.  The cost of removing the explicit
dependence on this physical energy scale is that the expansion will
therefore only be reliable for pion masses below this characteristic
scale $\mpi\lesssim 0.3$--$0.4\gev$.

We now describe the chiral expansion within finite-range
regularization, where the cut-off scale remains explicit. In
particular, we highlight the mathematical equivalence of FRR and
dimensional regularization in the low energy regime. We introduce a
functional cutoff, $u(k)$, defined such that the loop integral is
ultraviolet finite,
\begin{equation}
I_\pi = \frac{2}{\pi} \int_0^\infty dk\, \frac{k^4 u^2(k)}{k^2+\mpi^2} \, .
\label{eq:INpiFRR}
\end{equation}
To preserve the infrared behavior of the loop integral, the regulator
is defined to be unity as $k\to 0$.  For demonstrative purposes, we
choose a dipole regulator $u(k)=(1+k^2/\Lambda^2)^{-2}$, giving
\begin{equation}
I_\pi^{\rm DIP} =
\frac{\Lambda^5(\mpi^2+4\mpi\Lambda+\Lambda^2)}{16(\mpi+\Lambda)^4} 
\,\sim \frac{\Lambda^3}{16} - \frac{5\Lambda}{16}\mpi^2 
+ \mpi^3 - \frac{35}{16\Lambda} \mpi^4 + \ldots \,,
\label{eq:INpiDIP}
\end{equation}
The first few terms of the Taylor series expansion, as shown, provide
the relevant renormalisation of the low-energy terms.
The renormalized expansion in FRR is therefore precisely equivalent to
Eq.~(\ref{eq:MSrenLNA}) up to ${\cal O}(\mpi^3)$ where the leading
renormalized coefficients are given by
\begin{equation}
c_0 = a_0 + \chi_\pi \frac{\Lambda^3}{16} \, ,\quad
c_2 = a_2 - \chi_\pi \frac{5\Lambda}{16}  \, .
\label{LambdaFree}
\end{equation}
As $a_0$ and $a_2$ are fit parameters, the value $\Lambda$ takes is
irrelevant and plays no role in the expansion to the order one is
working; in this case $m_\pi^3$.  Retaining the full form of
Eq.~(\ref{eq:INpiDIP}) will therefore build in a resummation of
higher-order terms in the chiral series. 

It is straight forward to extend this procedure to next-to-leading
nonanalytic order, explicitly including all terms up to $m_q^2\sim
\mpi^4$.  Most importantly, there are nonanalytic contributions of
order $\mpi^4\log\mpi$ arising from the $\Delta$-baryon and tadpole
loop contributions.  The latter arises from the expansion of the
${\cal O}(m_q)$ chiral Lagrangian, with a coupling proportional to
the renormalized $m_q$ expansion coefficient, $c_2$.  In the heavy
baryon limit
\begin{equation}
\chi_{\pi\Delta} I_{\pi\Delta} =  - \frac{3}{16\pi^2\fpi^2} 
   \frac{16\, g_A^2}{9}
\int_0^\infty dk \frac{k^4\, u^2(k)}
{\omega(k) ( \Delta + \omega(k) )}\, ,
\label{eq:fullSEpi}
\end{equation}
\begin{equation}
\chi_{\rm tad} I_{\rm tad} = -\frac{3}{16\pi^2\fpi^2}\, c_2\, \mpi^2 
\left\{ \int_0^\infty dk 
\left(\frac{2 k^2\, u^2(k)}{\sqrt{k^2+\mpi^2}}\right) -
  t_0\right\}\, ,
\label{eq:tad}
\end{equation}
where $\omega(k)=\sqrt{k^2+\mpi^2}$ and $\Delta = 292$ MeV is the
physical $\Delta$-$N$ mass splitting.  The finite-range regulator
$u(k)$ is taken to be either a sharp theta-function cut-off, a dipole,
a monopole or finally a Gaussian.  These regulators have very
different shapes, with the only common feature being that they
suppress the integrand for momenta greater than $\Lambda$.  The
coefficient of Eq.~(\ref{eq:fullSEpi}) reproduces the empirical width
of the $\Delta$ resonance.

In Eq.~(\ref{eq:tad}), $t_0$ is defined such that the term in braces
vanishes at $\mpi=0$.  $t_0$ is a local counter term introduced in FRR
to ensure a linear relation for the renormalization of $c_2$.
This approach provides a relatively simple way to ensure that the
coefficient of the tadpole diagram contribution is proportional to the
renormalized coefficient, $c_2$, as opposed to a combination of
unrenormalized coefficients as done in Ref.~\cite{Bernard:2003rp}.
Nonanalytic terms of the chiral expansion must have renormalized
coefficients in order to recover the standard expansion of \chiPT\
upon taking the finite-range regulator parameter to infinity.

The key feature of finite-range regularization is the presence of an
additional adjustable regulator parameter which provides an
opportunity to suppress short distance physics from the loop integrals
of effective field theory.  As emphasized in Eq.~(\ref{eq:finite}) by
the superscripts $\Lambda$, the unrenormalized coefficients of the
analytic terms of the FRR expansion are regulator-parameter dependent.
However, the large $m_\pi$ behavior of the loop integrals and the
residual expansion (the sum of the $a_i^\Lambda$ terms) are remarkably
different.  Whereas the residual expansion will encounter a power
divergence, the FRR loop integrals will tend to zero as a power of
$\Lambda/m_\pi$, as $m_\pi$ becomes large.  Thus, $\Lambda$ provides
an opportunity to govern the convergence properties of the residual
expansion and thus the FRR chiral expansion.

Since hadron masses are observed to be smooth, almost linear functions
of $m_\pi^2$ for quark masses near and beyond the strange quark mass,
it should be possible to find values for the regulator-range
parameter, $\Lambda$, such that the coefficients $a_4^\Lambda$ and
higher are truly small.  In this case the convergence properties of
the residual expansion, and the loop expansion are excellent and their
truncation benign. Note that $\Lambda$ is not selected to approximate
the higher order terms of the chiral expansion.  These terms simply
sum to zero in the region of large quark mass and the details of
exactly how each of the terms enter the sum are largely irrelevant.

The FRR expansion is summarized in Eqs.~(\ref{eq:finite}),
(\ref{eq:chiNpi}), (\ref{eq:INpiFRR}), (\ref{eq:fullSEpi}) and
(\ref{eq:tad}).  Constraining the parameters $a_{0,2,4,6}^\Lambda$ to
lattice simulation results provides the extrapolations of
Fig.~\ref{FRR}.  Table~\ref{tab:cren} summarizes the associated
expansion coefficients obtained from various regulator fits to lattice
data.  Whereas uncertainties in Ref.~\cite{Leinweber:2003dg} were
reported at the 95\% confidence level, we report standard errors
herein.  The convergence of the residual expansion as represented by
the coefficients $a_{0,2,4}^\Lambda$ is excellent for the FRR
expansions and contrasts the coefficients of minimal subtraction as
represented by dimensional regularization.  Moreover, the level of
agreement among the renormalized quantities $M_N$, $c_0$ and $c_2$ for
the FRR results is remarkable, rendering systematic errors associated
with the shape of the regulator at the fraction of a percent level.

\begin{table}
\begin{center}
\caption{ Expansion coefficients and the corresponding extrapolated
nucleon mass obtained from various regulator fits to lattice data.
All quantities are in units of appropriate powers of GeV.  Errors are
statistical in origin arising from lattice data.  Deviations in the
central values indicate systematic errors associated with the chiral
extrapolation.
\label{tab:cren}}
\begin{tabular}{lccccccccc}
\hline\hline
                &\multicolumn{3}{c}{Bare Coefficients} &&\multicolumn{3}{c}{Renormalized Coefficients} \\
Regulator       & $a_0^\Lambda$    & $a_2^\Lambda$    & $a_4^\Lambda$    
                                                 &$\Lambda$ & $c_0$       & $c_2$       & $c_4$      & $M_N$         \\
\hline
Monopole	& $1.74$   & $0.30$   & $-0.49$  & $0.5$    & $0.923(65)$ & $2.45(34)$  & $20.5(15)$ & $0.960(58)$   \\
Dipole		& $1.30$   & $0.37$   & $-0.49$  & $0.8$    & $0.922(65)$ & $2.49(33)$  & $18.9(15)$ & $0.959(58)$   \\
Gaussian	& $1.17$   & $1.22$   & $-0.50$  & $0.6$    & $0.923(65)$ & $2.48(34)$  & $18.3(15)$ & $0.960(58)$   \\
Sharp cutoff	& $1.06$   & $0.56$   & $-0.55$  & $0.4$    & $0.923(65)$ & $2.61(33)$  & $11.8(13)$ & $0.961(58)$   \\
Dim.~Reg.	& $0.79$   & $4.15$   & $+8.92$  & --       & $0.875(56)$ & $3.14(25)$  & $7.2(8)$   & $0.923(51)$   \\
\hline\hline
\end{tabular}
\end{center}
\end{table}

In this light, it is interesting to consider the importance of
two-loop contributions to the chiral expansion \cite{McGovern:1998tm}.
The leading nonanalytic behavior is $m_\pi^5$ and $m_\pi^5 \log
m_\pi$.  However, we have already emphasized that the inclusion of
$m_\pi^5$ contributions leads only to small differences in the
extrapolation function, as illustrated in Fig.~\ref{FRR} and supported
by $M_N$, $c_0$ and $c_2$ of Table~\ref{tab:cren}.  Whereas the
coefficient of the $m_\pi^5$ term, $\chi_5$, vanishes for the
theta-function regulator, the monopole, dipole and Gaussian regulators
provide $\chi_5 = -44$, $-34$ and $-30\gev^{-4}$ respectively.  The
point is that by the time the pion mass is sufficiently large to allow
this term to contribute significantly, other complementary terms
conspire to make the sum of terms of the FRR loop integral small.
Indeed some interplay between $c_4$ and $\chi_5$ is observed in
Table~\ref{tab:cren}, where $c_4$ displays a correlation with the
magnitude of $\chi_5$.
Terms at fifth order will reveal themselves in precision effective
field theory studies well within the power-counting regime of \chiPT,
where $m_\pi^6$ contributions are small relative to the $m_\pi^5$ and
$m_\pi^5 \log m_\pi$ terms of interest.  Again, at larger pion masses,
all higher-order terms become large and given the lattice QCD results,
these higher-order terms must sum to zero.

The uncertainty associated with the coefficient of $\mpi^5$ has also
been highlighted by Beane in Ref.~\cite{Beane:2004ks}. The traditional
chiral expansion was demonstrated to show a strong sensitivity to the
value of this coefficient.  By accepting a 20\% error associated with
the truncation of the series, he suggested that the power counting
regime might be restricted to within $\mpi\lesssim 300\mev$.  This is
supported by Table~\ref{tab:cren} where one can see that the
renormalized low-energy coefficients of the expansion, $c_i$, have
been sacrificed in applying the truncated expansion of \chiPT\ to the
range $0 \le m_\pi^2 \le 1.0\gev^2$.

\begin{figure}[t]
\begin{center}
\includegraphics[width=0.9\hsize]{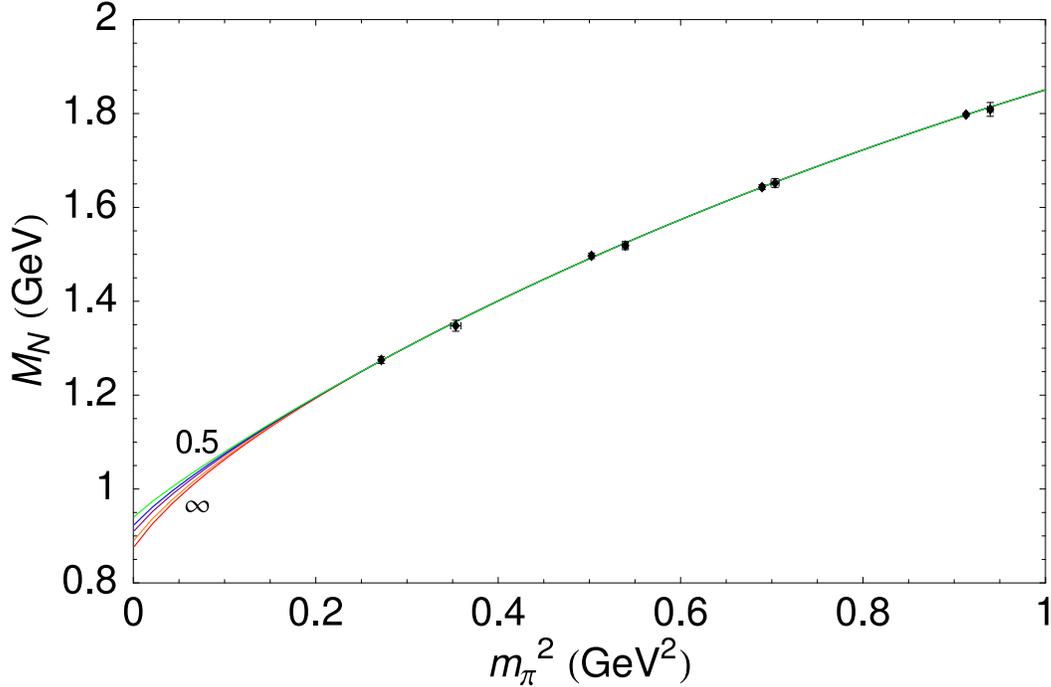}
\end{center}
\vspace{-1.3cm}
\caption{Dipole regulator fits to lattice data for various dipole
regularization scales, $\Lambda=0.5,\ 0.8,\ 1.1,\ 2.0,\ {\rm and}\
\infty\gev$, interpolating between FRR and minimal subtraction.  
\vspace{-0.3cm}
}
\label{FRRtoMS}
\end{figure}

Figure~\ref{FRRtoMS} further emphasizes the need to know the power
counting regime of \chiPT.  As one moves the regulator parameter away
from the regime of 1 GeV, the otherwise good convergence of the
residual expansion is lost, and the truncation of the expansion is no
longer benign.  The naive application of the minimal subtraction
scheme outside the power-counting regime leads to a steeper slope in
the nucleon mass relevant to the pi-nucleon sigma term, and produces
an artificially large value.

\section{POWER-COUNTING REGIME}

Because the chiral expansion of \chiPT\ is truncated with no attempt
to estimate the contribution of higher-order terms, knowing the
power-counting regime of a given truncation is absolutely essential to
extracting correct physics from \chiPT.  One simply obtains the wrong
answer if one works outside the power-counting regime.

Now that the renormalized low-energy coefficients have been determined
to the order we are working, we can use this information to estimate
the power-counting regime of \chiPT.  In the following, we quantify
the power-counting regime by identifying the range of pion masses
whereby the expansion is truly independent of regularization scheme
and hence independent of contributions from higher-order terms in the
expansion.  As discussed in the introduction and in detail in
Sec.~\ref{renormalization} surrounding Eq.~(\ref{LambdaFree}), the
FRR chiral expansion is mathematically equivalent to that of \chiPT\
to the finite order one is working.  In other words, these terms of
the FRR expansion are independent of the regulator parameter, and any
value of $\Lambda$ is allowed.  

\begin{figure}[!t]
\begin{center}
\includegraphics[width=7.9cm]{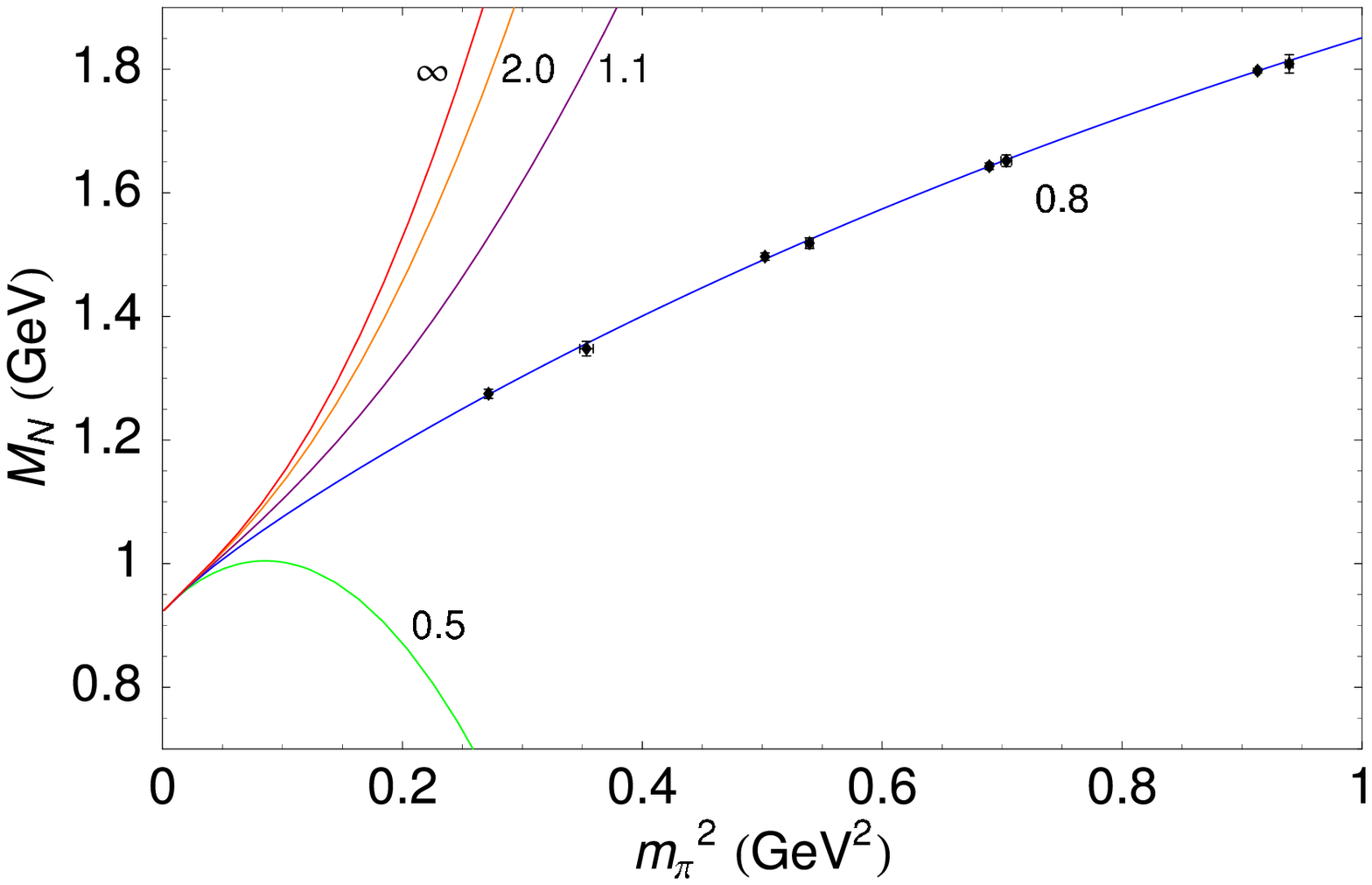}
\includegraphics[width=7.9cm]{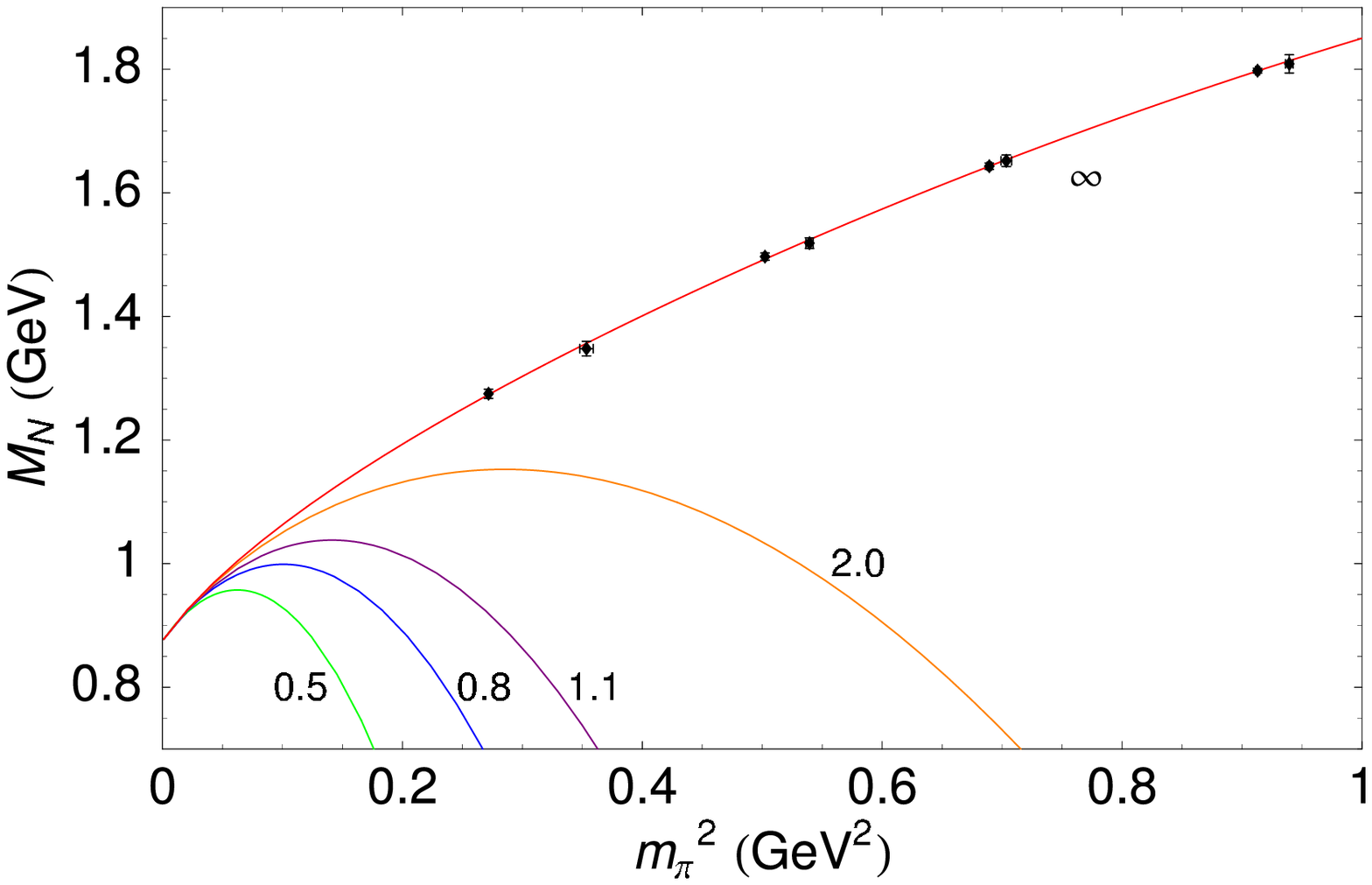}
\end{center}
\vspace{-1.3cm}
\caption{With the low-energy parameters $c_0$, $c_2$ and $c_4$ fixed
to those obtained by the fit to lattice data, the chiral expansion is
shown for various values of the dipole regulator scale, $\Lambda=0.5,\
0.8,\ 1.1,\ 2.0\ {\rm and}\ \infty\gev$.  The left panel corresponds
to the parameters $c_i$ determined by the $\Lambda=0.8\gev$ dipole
regulator fit and the right panel to those determined in the 
$\Lambda\to\infty$ minimal subtraction limit.
\vspace{-0.3cm}
}
\label{fig:fixC}
\end{figure}

Fig.~\ref{fig:fixC} illustrates the fourth-order chiral expansion for
various dipole regulator parameters $\Lambda$, ranging between
$0.5\gev$ and $\infty$.  This therefore describes a continuous
transition between finite-range regularization and a minimal
subtraction scheme.  The renormalized coefficients and nonanalytic
terms to fourth order are automatically independent of $\Lambda$.  The
observed changes in the curves are simply a reflection of the changes
in terms beyond the order one is working.  Of course, only the curves
with $\Lambda \sim 1$ GeV have a residual expansion with good
convergence properties allowing truncation of the expansion, without a
loss of accuracy.

\begin{figure}[t]
\begin{center}
\includegraphics[width=0.9\hsize]{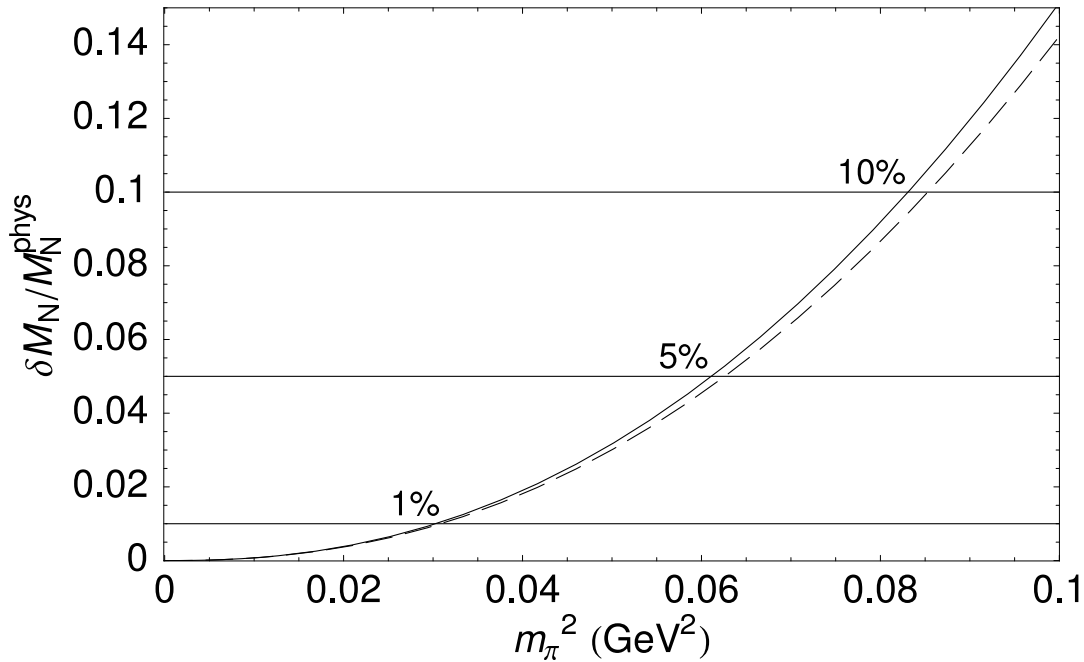}
\end{center}
\vspace{-1.3cm}
\caption{For fixed low-energy coefficients $c_0$, $c_2$ and $c_4$, the
relative difference in the nucleon mass expansion for two extremal
regularization scales, where $\delta
M_N=M_N(\Lambda=\infty)-M_N(\Lambda=0.5)$.  The solid and dashed curve
corresponds to the values of $c_i$ determined from the fit to lattice
results with dipole of scale $\Lambda=0.8$ and $\infty\gev$,
respectively.
\vspace{-0.3cm}
}
\label{fig:pcregime}
\end{figure}

The power counting regime can be identified as the regime where the
curves are independent of $\Lambda$ at some level of precision.  As
discussed earlier, this must be at the fraction of a percent level in
order to predict the nucleon mass to an accuracy of one percent.  The
regime within which the curves agree within one percent is $0 \le
m_\pi \le 0.18\gev$. This is displayed in Fig.~\ref{fig:pcregime},
where the relative error between the two extremal regularization
scales shows the sensitivity to higher orders in the chiral expansion.

This sensitivity to higher-order terms also indicates that tolerance
levels of 5\% and 10\% can be achieved up to pion masses of $0.25$ and
$0.29\gev$, respectively.  These estimates of the chiral regime are
similar to those reported by Beane \cite{Beane:2004ks}, as discussed
above.

The results of Fig.~\ref{fig:pcregime} are robust to the choice of
$\Lambda$ used to obtain the leading renormalized low-energy
coefficients $c_0$, $c_2$ and $c_4$ of the chiral expansion.
Figure~\ref{fig:fixC} also displays results in the extreme (and
undesirable) case where the coefficients are fixed in the
$\Lambda\to\infty$ (minimal subtraction) limit.  The associated
power-counting regime is displayed in Fig.~\ref{fig:pcregime} by the
dashed curve, revealing a negligible difference.

\section{CONCLUSIONS}

As we have strictly shown, FRR is mathematically equivalent to minimal
subtraction schemes such as dimensional regularization to any finite
order one wishes to work.  When working within the power-counting
regime as defined above, the FRR and minimal-subtraction schemes are
model independent.  

However, when used beyond the power-counting regime, the truncated
expansion of \chiPT\ fails.  The signature of this failure is that the
low-energy coefficients of the expansion are sacrificed to fit the
lattice results, leading to values that differ from the correct values
of the low-energy theory.  Knowledge of the extent of the
power-counting regime is as important as knowledge of the finite-order
chiral expansion itself.

Methods to determine the power counting regime have been described
herein.  If one limits the systematic errors at the fraction of a
percent level to allow a prediction of the nucleon mass at the 1\%
level, the power-counting regime is 
$0 \le m_\pi^2 \le 0.03\gev^2$ or $0 \le m_\pi \le 0.18\gev$.
Unfortunately, today's lattice QCD results do not come close to this
power-counting regime of fourth-order \chiPT.  With the regime
extending marginally beyond the physical pion mass, one might wonder
what role, if any, traditional \chiPT\ will play in the extrapolation
of lattice QCD results.  The problem is exacerbated by the need to
have sufficient data inside the power-counting regime to determine the
fit parameters.  In time, algorithms which enable
dynamical-fermion simulations below
$m_\pi = 0.18\gev$,
are likely to permit a direct simulation at $m_\pi = 0.140$ GeV.

In contrast, FRR schemes take on a model dependence when used outside
the power-counting regime.  The model-dependence associated with the
shape of the regulator is at the fraction of a percent level, even
over the range $0 \le m_\pi^2 \le 1.0\gev^2$.  As the expansion is
truncated there is a residual dependence on the regulator parameter,
$\Lambda$, which can be quantified by varying $\Lambda$ over a range
where convergence of the residual expansion remains acceptable.  The
success of FRR is realized in that this residual degree of
model-dependence in demonstrably small, such that the chiral
extrapolation of modern lattice results is under control.
This remarkable result is extremely important in that it is likely to
provide the only reliable method to extract physical hadron properties
from lattice simulation results for many years.


DBL thanks the Theory Group at Jefferson Lab for their kind
hospitality during his stay while on study leave.  He also thanks Tom
Cohen, Stefan Scherer, Jambul Gegelia, Matthias Schindler and Vladimir
Pascalutsa for interesting and helpful discussions.  This work was
supported by the Australian Research Council and by DOE contract
DE-AC05-84ER40150, under which SURA operates Jefferson Laboratory.

\end{document}